\documentclass[a4paper,11pt]{article}
\usepackage{amsmath}
\usepackage{amsbsy}
\usepackage{amsthm}
\usepackage{amsfonts}
\usepackage{amssymb}
\usepackage{graphicx}
\usepackage{epstopdf}
\usepackage{tikz}
\usepackage{dsfont}
\usepackage{authblk}

\newcommand{\bs}[1]{{\ensuremath{\boldsymbol{#1}}}}

\newtheorem{theorem}{Theorem}

\begin{document}

\title{Conformal invariance and Renormalization Group}

\author[1,2]{Alessandro Giuliani}
\affil[1]{\small{Universit\`a degli Studi Roma Tre, Dipartimento di Matematica e Fisica, L.go S. L. Murialdo 1, 00146 Roma, Italy}}
\affil[2]{\small{Centro Linceo Interdisciplinare {\it Beniamino Segre}, Accademia Nazionale dei Lincei, Palazzo Corsini, Via della Lungara 10,
00165 Roma, Italy.}}

\date{}

\maketitle

\begin{abstract}
Conformal field theory (CFT) is an extremely powerful tool for explicitly computing critical exponents and correlation functions of statistical mechanics systems at a second order phase transition, or of condensed matter systems at a quantum critical point. Conformal invariance is expected to be a feature of the fixed point theory obtained from a microscopic model at criticality, under appropriate averaging and rescaling operations: the action of the Wilsonian Renormalization Group (RG). Unfortunately, an explicit connection between critical microscopic models and their conformally invariant scaling limit is still lacking in general. Nevertheless, the last decades witnessed significant progress on this topic, both from the mathematical and physics sides, where several new tools have been introduced and their ranges of applications have constantly and significantly increased: I refer here, e.g., to discrete holomorphicity, SLE, the use of lattice Ward Identities in constructive RG, the conformal bootstrap program and its recent applications to 3D CFT. In an effort to make further progress on these problems, the one-day workshop {\it Emergent CFTs in statistical mechanics} was organized and held at Institut Curie in Paris on January 29, 2020: the goal was to bring together probabilists, mathematical physicists and theoretical physicists, working on various aspects of critical statistical mechanics systems with complementary tools, both at the discrete and the continuum level, in the hope of creating new connections between the different approaches. This paper is based on an introductory talk given at the workshop: after a summary of the main topics discussed in the meeting, I illustrate the approach to the problem based on constructive RG methods, by reviewing recent results on the existence and the explicit characterization of the scaling limit of critical 
2D Ising models with finite range interactions in cylindrical geometry. 
\end{abstract}

\section{Introduction}
The general questions motivating today's workshop\footnote{{\it Emergent CFTs in statistical mechanics}, part of the series {\it Inhomogeneous Random Systems} (organizers: Fran\c{c}ois Dunlop and Ellen Saada; moderator: Alessandro Giuliani) held at Institut Curie in Paris on January 29, 2020. 
This paper is based on an introductory talk given at this workshop.} are the following: given a lattice statistical mechanics model at a second order phase transition point, how do we prove that the system admits a scaling limit? How do we prove that the limit, when it exists, is conformally invariant? And how do we 
explicitly identify it? 
 
The very existence and conformal invariance of the scaling limit of critical statistical mechanics systems is strongly suggested by Wilsonian Renormalization Group (RG) \cite{W1,W2,W3}: in this framework, scaling limits correspond to fixed points of the RG transformation; universality classes are understood in terms of basin of attractions of such fixed points;
microscopic lattice Hamiltonians are special initial data in the space of Hamiltonians, which the RG transformation acts on. 
Scale invariance of the fixed point follows automatically from the Wilsonian construction. Under a few additional assumptions, expected to be valid in great generality for local models, scale invariance is promoted to {\it conformal} invariance, as first discussed by Polyakov \cite{P70} and later by Zamolodchikov \cite{Z86} and Polchinski \cite{Po88}, among others.

Even though the previous scheme is generally believed to be {\it the} justification for the existence and conformal invariance of the scaling limit, there are very few 
cases for which one can {\it prove} (mathematically -- or, at least, via a systematic argument that does not throw away a priori terms that `are expected to be negligible', 
without a way to compute or estimate them) existence of the limit and {\it identify} it with the appropriate Conformal Field Theory (CFT). 

The last 20 years witnessed remarkable progress on the understanding of these questions, both on the mathematical and on the theoretical physics sides, which 
allowed to exhibit the first examples of conformally invariant scaling limits, rigorously constructed starting from lattice microscopic models, as well as to 
predict the critical exponents of several critical statistical mechanics systems in three or more dimensions, at a better precision than via MonteCarlo simulations. The 
speakers of today's meeting belong to three different areas, which contributed substantially to these developments from complementary perspectives. Let me briefly introduce these areas and the corresponding speakers. 

\begin{enumerate}
\item {\it Probability, Geometry of random curves and Discrete holomorphicity. } This is the area which 
Federico Camia, Cl\'ement Hongler and Dmitry Chelkak belong to. The approach based on these methods led to the 
complete proof of conformal invariance of the scaling limit of the two dimensional nearest neighbor Ising \cite{CHI15,CS09,DS12,HS13,Smi10} 
and dimer models \cite{K00,K01}. It has the 
advantage of being flexible in treating geometric deformations of the domain and of the underlying lattice, thus leading to the first proofs of universality 
with respect to these kinds of deformations. The limitation of this approach is that it is mostly\footnote{There are a few notable exceptions: I refer to the recent
results obtained by this approach on non-exactly solved models and/or models away from the free Fermi point and, more specifically on: crossing probabilities for critical percolation on triangular lattice \cite{Smi01}; Pfaffian nature of boundary spin correlations in interacting Ising models \cite{ADTW};  limit shapes and surface tension for the 5V model \cite{dGKW}; height (de)localization transition in the 6V model \cite{GP}.}
restricted to exactly solved models at the free Fermi point, 
such as nearest neighbor Ising and dimers in two dimensions, and it is not flexible 
in dealing with perturbations of the microscopic Hamiltonian. 
\item {\it Constructive RG.} This is the area which I and Vieri Mastropietro belong to. The approach based on these methods led to the construction of the bulk scaling limit of several interacting, non-solvable, models, such as $\varphi^4_4$ \cite{BBS,GK}, $\varphi^4_3$ with long-range interactions \cite{ACG13,BDH,BMS}, 
sine-Gordon on the Kosterlitz-Thouless critical line \cite{F}, 1D lattice interacting fermions \cite{BGPS}, 1D quantum spin chains with finite range interactions \cite{BM01}, 2D interacting dimers and 6-vertex models \cite{GMT20}, Ashkin-Teller and 8-vertex models \cite{BFM09b,GM05}, 2D graphene \cite{GM10}, 3D Weyl semimetals \cite{GMP20,M14}, and many others. Remarkably, 
many of the models listed here have non-Gaussian or non-determinantal infrared fixed points, and results are robust under a large class of microscopic perturbations 
of the lattice Hamiltonian. Moreover, this approach led to the proof of several predictions from CFT, such as scaling relations among critical exponents and amplitudes
\cite{BM11,GMT20}, 
bosonization identities \cite{BW20,BFM09a}, and expression for the universal subleading contributions to the critical free energy \cite{GM13}. A limitation of this approach is that it is 
restricted to `weakly interacting' cases, that is, to models that are close to a Gaussian model or to a free Fermi model. Moreover, it is not flexible in dealing with geometric perturbations of the domain and/or of the underlying lattice: more generally, it is mostly\footnote{A couple of exceptions are the results of \cite{ACG13} on a $p$-adic hierarchial version of $\varphi^4_3$ with long-range interactions (see also \cite{A18} for a review on the subject, 
trying to establish a bridge between RG and CFT, very much in the spirit of the present paper), 
and the results of \cite{AGG20} on the scaling limit of 2D Ising models with finite range interactions in cylindrical geometry, reviewed below, in Section \ref{sec:2}).} limited to translationally invariant situations.

\item {\it CFT and Conformal bootstrap.} This is the area which Jesper Jacobsen and Slava Rychkov belong to. 
These methods led to exact solutions of several CFTs, in particular exact predictions for critical exponents and closed formulas (or closed equations) for correlation functions of any order. In two dimensions, essentially all possible CFTs have been identified and solved \cite{BPZ84}, with 
very precise explicit predictions on the spectrum of critical exponents and structure of the correlation functions. In three or more dimensions, the constraints from the conformal bootstrap, in combination with numerics, provided rigorous bounds on the critical exponents of several strongly interacting models, 
most notably the 3D Ising model \cite{PRV19}; predictions are in some cases more precise than the best numerical MonteCarlo estimates. Potentially, the conformal bootstrap program could lead to exact solutions of non-Gaussian CFTs in three or more dimensions. 
A limitation of this approach is that it is axiomatic: its predictions rely on a number of assumptions (conformal covariance, Operator Product Expansion, ...)
that are very hard to prove (if not impossible at the present state-of-the-art) starting from microscopic models. The identification of a given microscopic model with 
its universality class is usually done indirectly (and non-rigorously), via symmetry considerations or by using a priori constraints on critical exponents, possibly 
following from other methods. 
\end{enumerate}
So far, these three communities did not talk to each other enough, even though I believe that progress will come from a better exchange of ideas among them. I hope that from today's workshop concrete proposals for connections among complementary approaches will emerge\footnote{An output of the constructive dialog that emerged from the workshop is the work \cite{GMR20}.}. 
A few natural questions and problems that could be attacked by a constructive dialog between these areas are the following: 
\begin{enumerate}
\item\label{it:1} Can the probabilistic approach, which is very flexible in dealing with scaling limits in non-trivial geometries, be combined with constructive RG techniques, which 
are very robust under irrelevant perturbations of the microscopic Hamiltonian, in order to construct the scaling limits of non-solvable models close to the 
free Fermi point in domains of arbitrary shape, and prove their conformal covariance? 
\item Can conformal perturbation theory be given a constructive meaning? Can it be interpreted or reformulated as an instance (or an extension) of constructive RG 
techniques in the vicinity of a non-trivial fixed point? 
\item Can one use constructive RG to substantiate some of the CFT axioms, such as the Operator Product Expansion? 
\end{enumerate}
Let me stress that, while there is certainly room for `local' progress, there are to date very big questions and open problems for which a strategy is 
completely missing, and for which the development of inter-disciplinary approaches would be even more urgent: for example, how can we construct 
very non-Gaussian fixed points via RG methods? Can ideas from conformal perturbation theory in combination with constructive RG be useful in this respect? 
Can informations from the exact or numerical solutions of hierarchical models with very non-Gaussian fixed points 
be exported to the realm of short-range, translationally invariant, models with an explicit  control on the error (and a systematic way to improve it)? 

\section{An illustrative example: the scaling limit of non-integrable 2D Ising models}\label{sec:2}

As mentioned above, I belong to the area of constructive RG, and I would like to illustrate some of the results we obtained via these methods for
critical 2D Ising models with finite-range interactions. In short, we succeeded in constructing the scaling limit of the multipoint energy correlations 
for a class of non-integrable Ising models in the full plane \cite{GGM12} and in cylindrical geometry \cite{AGG20}. 
Extensions to more general domains and proof of conformal covariance of the limit
will presumably require additional inputs from probabilistic and discrete holomorphicity methods, in the spirit of problem/question \ref{it:1} in the list at the end of the 
previous section. In the following I will describe the setting, state more precisely our main results, and give a sketch of the proof. 

\medskip

Consider a finite rectangular portion $\Lambda^a_{L,M}\equiv\Lambda$ of $a\mathbb Z^2$ ($a$ being the lattice spacing) of horizontal 
side $\ell_1=aL$ and vertical side $\ell_2=aM$, with $L,M$ two integers (in other words, the rectangle consists of $L$ columns and $M$ rows) centered at the origin. 
We are interested in two types of boundary conditions: either periodic in both horizontal and vertical direction, or periodic in the horizontal and free in the vertical direction. In the first case, $\Lambda$ is a discrete torus, in the second it is a discrete cylinder. 

We consider the following Hamiltonian:
\begin{equation}H_{\Lambda}= -J\sum_{\langle x,y\rangle}\sigma_x\sigma_y+\lambda\sum_{X\subset \Lambda} V(X)\sigma_X\equiv H^{0}_\Lambda+\lambda V_\Lambda,\end{equation}
where: $\sigma_x=\pm1$ are Ising spins; the first sum runs over (unordered) nearest neighbor pairs of $\Lambda$; in the second sum, given a subset $X$ of $\Lambda$, we denoted 
$\sigma_X=\prod_{x\in X}\sigma_x$ and $V(X)$ is a finite range, translationally invariant interaction, supported on {\it even} sets $X$.
For example, by choosing $V(X)$ appropriately, the term $\lambda V_\Lambda$ reduces to the pair interaction 
$\lambda\sum_{\langle\!\langle x,y\rangle\!\rangle}\sigma_x\sigma_y$ 
where the sum runs over pairs of next-to-nearest-neighbor sites; we remark that no particular simplification in the proofs takes place in this case. 

Let $\Lambda_{\ell_1,\ell_2}$ be the continuous torus or cylinder of sides $\ell_1$, $\ell_2$, centered at the origin. Given any $x$ in the interior of $\Lambda_{\ell_1,\ell_2}$, 
we let 
\begin{equation}\varepsilon^a_j(x)=a^{-1}\big(\sigma_{[x]}\sigma_{[x]+a\hat e_j}-\langle \sigma_{[x]}\sigma_{[x]+a\hat e_j}
\rangle^\lambda_{\beta_c(\lambda);\Lambda}\big),\end{equation} where $[x]$ denotes the closest point to $x$ among those of $\Lambda$ (in case of ambiguity, we choose the closest to the left/bottom of $x$), $\hat e_j$ is the unit coordinate vector in direction $j\in\{1,2\}$, 
$\langle \cdot \rangle^\lambda_{\beta;\Lambda}$ is the Gibbs measure with weight proportional to $e^{-\beta H_\Lambda}$ (the label $\lambda$ is meant to 
emphasize the fact that the measure depends on the interaction of strength $\lambda$), and $\beta_c(\lambda)$ is the critical temperature, still to be determined 
(and that, in general, is expected to depend on $\lambda$; for $\lambda=0$ it is well known to be $\beta_c(0)=\tanh^{-1}(\sqrt2-1)$). 
We are interested in the multipoint energy correlations
\begin{equation} \langle \varepsilon^a_{j_1}(x_1)\cdots \varepsilon^a_{j_n}(x_n)\rangle^\lambda_{\beta_c(\lambda);\Lambda}\end{equation}
for $n\ge 2$, in the limit $a\to 0$ and $L,M\to\infty$. There are two natural ways of performing these limits, and we shall be concerned with both: either we send 
$L,M\to\infty$ first and then $a\to 0$, or we perform the limits simultaneously, in such a way that $aL\to \ell_1$ and $aM\to\ell_2$, with $\ell_1,\ell_2$ two positive real 
numbers. The first case will lead to the computation of the scaling limit in the infinite plane, while the second to the scaling limit in the finite torus or cylinder 
$\Lambda_{\ell_1,\ell_2}$, depending on the considered boundary conditions. 
The critical temperature $\beta_c(\lambda)$ is fixed in such a way that\footnote{In eq.\eqref{eq:4} and below $\lim_{\Lambda\nearrow a\mathbb Z^2}$ denotes the limit 
$L,M\to\infty$, performed in such a way that $C^{-1}\le L/M\le C$ for some $C>0$.}
\begin{equation} \lim_{\Lambda\nearrow a\mathbb \mathbb Z^2}\langle \varepsilon^a_{j_1}(x_1) \varepsilon^a_{j_2}(x_2)\rangle^\lambda_{\beta_c(\lambda);\Lambda}
\label{eq:4}\end{equation}
decays polynomially (rather than exponentially) to zero as $|x_1-x_2|\to\infty$. $\beta_c(\lambda)$ is known to be well-defined and unique for $\lambda V_\Lambda$ 
a ferromagnetic pair interaction, via FKG correlation inequalities. In more general cases, uniqueness of $\beta_c(\lambda)$ follows from the proof underlying the results stated below, provided $\lambda$ is small enough. 

\subsection{Main results}

Our main results concern the scaling limit of the multipoint energy correlations in the infinite plane and in the cylinder of sides $\ell_1,\ell_2$. The result for the infinite plane can be formulated as follows: 
\begin{theorem} \cite{GGM12} Let $\Lambda=\Lambda^a_{L,M}$ be the discrete torus introduced above and $V$ an even, finite range, translationally invariant interaction. For $\lambda$ small enough, there exist two real analytic functions $\beta_c(\lambda)$ and $Z(\lambda)$, such that $\beta_c(0)=\tanh^{-1}(\sqrt2-1)$, $Z(0)=1$ and
\begin{equation}\lim_{a\to 0}\lim_{\Lambda\nearrow a\mathbb Z^2} \langle \varepsilon^a_{j_1}(x_1)\cdots \varepsilon^a_{j_n}(x_n)\rangle^\lambda_{\beta_c(\lambda);\Lambda}=(Z(\lambda))^n\Big(\frac{i}{\pi}\Big)^{n}{\rm Pf} K_{\mathbb R^2}(x_1,\ldots,x_n),\end{equation}
where ${\rm Pf}$ denotes the Pfaffian and $K_{\mathbb R^2}(x_1,\ldots,x_n)$ is the anti-symmetric matrix with elements $\big(K_{\mathbb R^2}(x_1,\ldots,x_n)\big)_{ij}=\frac{\mathds 1_{i\neq j}}{z_i-z_j}$, with $z_j=(x_j)_1+i(x_j)_2$ the complex 
representative of $x_j$. \label{thm:1}
\end{theorem}
As discussed in the following, the proof of this theorem relies (too heavily) on the translation invariance of the model. Breaking translation invariance leads to new difficulties that, for the moment, we managed to overcome in the case of cylindrical geometry, in which case we obtain the following: 
\begin{theorem} \cite{AGG20} Let $\Lambda=\Lambda^a_{L,M}$ be the discrete cylinder introduced above and $V$ an even, finite range, translationally invariant interaction. Let $\ell_1$ and $\ell_2$ be two positive real numbers such that $C^{-1}\le \ell_1/\ell_2\le C$ for some positive constant $C$. 
For $\lambda$ small enough, the same $\beta_c(\lambda), Z(\lambda)$ as in the previous theorem, and any $n$-ple of points $x_1,\ldots, x_n$ in the interior of $\Lambda_{\ell_1,\ell_2}$ with $n\ge 2$, 
\begin{eqnarray}&& \lim_{\substack{a\to 0,\ L,M\to \infty\,: \\ aL\to \ell_1,\ aM\to\ell_2}}\ \langle \varepsilon^a_{j_1}(x_1)\cdots \varepsilon^a_{j_n}(x_n)\rangle^\lambda_{\beta_c(\lambda);\Lambda}=\\
&&\qquad =(Z(\lambda))^n\lim_{\substack{a\to 0,\ L,M\to \infty\,: \\ aL\to \ell_1,\ aM\to\ell_2}}\ \langle \varepsilon^a_{j_1}(x_1)\cdots \varepsilon^a_{j_n}(x_n)\rangle^0_{\beta_c(0);\Lambda}.\nonumber\end{eqnarray}
[Note that $\lambda=0$ in the right side of this equation.] The limit in the  right side can be rewritten as
\begin{equation}\lim_{\substack{a\to 0,\ L,M\to \infty\,: \\ aL\to \ell_1,\ aM\to\ell_2}}\ \langle \varepsilon^a_{j_1}(x_1)\cdots \varepsilon^a_{j_n}(x_n)\rangle^0_{\beta_c(0);\Lambda}=\Big(\frac{i}\pi\Big)^n{\rm Pf} K_{\Lambda_{\ell_1,\ell_2}}(x_1,\ldots,x_n),\end{equation}
for a suitable anti-symmetric matrix $K_{\Lambda_{\ell_1,\ell_2}}(x_1,\ldots,x_n)$.
\label{thm:2}
\end{theorem}

{\bf Remarks.} 
\begin{enumerate}
\item The matrix $K_{\Lambda_{\ell_1,\ell_2}}(x_1,\ldots,x_n)$ has an explicit expression, analogous to the one in the infinite plane (see Theorem \ref{thm:1}), which can be obtained from $K_{\mathbb R^2}(x_1,\ldots,x_n)$ by replacing the `Dirac propagator' $\frac1{z_i-z_j}$ 
by its counterpart on the cylinder of sides $\ell_1,\ell_2$ (whose definition involves an appropriate `image rule', see \cite{AGG20}). 
\item Theorem \ref{thm:2} is uniform in $\ell_1,\ell_2$, provided $\ell_1/\ell_2$ is bounded from above and below, as stated in the assumptions of the theorem. In particular, we 
can take $\ell_1,\ell_2\to\infty$ after having re-centered the cylinder at the point of coordinates $(0,\ell_2/2)$, 
in which case the multipoint energy correlations tend to those in the half-plane $\mathbb H$.
\end{enumerate}

The main point of Theorem \ref{thm:2}, as compared with Theorem \ref{thm:1}, is the presence of a boundary. The generalization has interest by itself, in that: (1)
the result is scale-covariant under changes of the aspect ratio; (2) it justifies the expected structure of the allowed boundary conditions in the scaling limit; (3)
it can be extended to boundary correlations, such as boundary spin and boundary energy correlations \cite{CavaPhD}. From my perspective, the result is interesting also because the proof of Theorem \ref{thm:2} requires to understand how to implement constructive RG in a non-translationally invariant setting, which is not technically well developed yet, and is very interesting for other related contexts, such as: boundary correlation and critical exponents; effect of defects and impurities (such as in the Kondo problem); effect of disorder and interactions (such as in Many Body Localization); effects of `cuts' with monodromy (such as those 
arising in the computation of spin-spin correlations in the Ising model, or monomer-monomer correlations in the dimer model). The methods introduced 
in the proof of Theorem \ref{thm:2}, combined with those used to construct the scaling limit of the nearest neighbor Ising model in domains of arbitrary shape, 
may lead, in perspective, to the proof of universality of the scaling with respect both to weak perturbations of the microscopic Hamiltonian and to 
geometric deformations of the domain and of the underlying lattice. 

\subsection{Sketch of the proof}

The proofs of Theorem \ref{thm:1} and \ref{thm:2} are based on the following strategy (note: formulas are simplified or approximate in order to convey the message with 
technical complications reduced to a minimum, see \cite{AGG20,GGM12} for additional details).  

\medskip

{\it 1. Grassmann representation.} The first step of the proof consists in deriving a representation of the partition function and of the generating function of energy correlations in terms of a Grassmann integral, with the structure 
of a fermionic $\lambda\psi^4_2$ theory. In particular, for all $\beta$, the partition function $Q_\Lambda=Q_\Lambda(\beta)$ can be schematically written as (similar formulas hold for the generating function of 
energy correlations): 
\begin{equation} Q_\Lambda= \int\mathcal D\psi\int \mathcal D\chi\, e^{-\frac12(\psi, C_c\psi)-\frac12(\chi, C_m\chi)+V(\psi,\chi)}.\label{eq:8} \end{equation}
Here $\psi=\{\psi_{\omega,x}\}^{\omega\in\{+,-\}}_{x\in\Lambda}$ and $\chi=\{\chi_{\omega,x}\}^{\omega\in\{+,-\}}_{x\in\Lambda}$ are two sets of Grassmann fields, and the symbols $\int \mathcal D\psi$, $\int \mathcal D\chi$ indicate the corresponding Grassmann (or Berezin) integrals. The terms $-\frac12(\psi, C_c\psi)$ and 
$-\frac12(\chi, C_m\chi)$ at exponent are the quadratic contributions to the bare Grassmann action and $V(\psi,\chi)$ is the interaction, of strength $\lambda$, and 
equal to the sum of Grassmann monomials in $\psi,\chi$ of order $2$, $4$, $6$, etc., whose kernels are analytic in $\lambda$ in a small neighborhood of the origin and  
decay exponentially to zero at large distances, with rate proportional to the inverse lattice spacing. The nearest neighbor Ising model corresponds to the case 
$\lambda=0$, in which case the interaction term $V$ vanishes: therefore, the partition function of the nearest neighbor model reduces to a Gaussian 
Grassmann, which can be computed explicitly in terms of Pfaffians or determinants (as well known). The matrices $C_c=C_c(\beta)$, $C_m=C_m(\beta)$ of the quadratic forms at exponent play the role of inverse covariance matrices of the $\psi$ and $\chi$ fields, respectively, for this Gaussian 
reference model (the nearest neighbor, `non-interacting', model). 
The labels  `$c$' and `$m$' standing for `critical' and `massive': these names are motivated by the fact that: (1) the {\it propagator} $g_m=C_m^{-1}$ has elements 
\begin{equation} [g_m(x,y)]_{\omega,\omega'}=\frac1{{\rm Pf}C_m}\int \mathcal D\chi e^{-\frac12(\chi,C_m\chi)}\chi_{\omega,x}\chi_{\omega',y}\end{equation}
decaying exponentially on the lattice scale, i.e., $\|g_m(x,y)\|$ is bounded by (const.)$a^{-1}e^{-\kappa |x-y|/a}$ for some constant $\kappa>0$, uniformly in the temperature $\beta$ and in the lattice scale; (2) the propagator $g_c=C_c^{-1}$ has elements 
\begin{equation} [g_c(x,y)]_{\omega,\omega'}=\frac1{{\rm Pf}C_c}\int \mathcal D\psi e^{-\frac12(\chi,C_c\chi)}\psi_{\omega,x}\psi_{\omega',y}\end{equation}
decaying polynomially at $\beta=\beta_c(0)$ and, more precisely, at that temperature $\|g_c(x,y)\|$ behaves asymptotically as (const.)$|x-y|^{-1}$ as $|x-y|\to\infty$. 
For $\beta\neq \beta_c(0)$, the elements of $g_c$ decay exponentially to zero, with rate going to zero as $\beta\to\beta_c(0)$ at speed 
$\propto a^{-1}(\beta-\beta_c(0))$. In order to avoid confusion, we will denote by $g_c^*$ the critical propagator computed at $\beta_c(0)$. 

\medskip

{\it 2. Integration of the massive field.} Thanks to the exponential decay of its propagator, the $\chi$ field can be integrated out in a `single shot', via a Grassmann 
version of the cluster expansion, based on the Battle-Brydges-Federbush-Kennedy formula \cite{BF, BF78, BK87}. The outcome is, letting $P(\mathcal D\psi)=({\rm Pf}C_c)^{-1}\mathcal D\psi
e^{-\frac12(\psi, C_c\psi)}$ be the Gaussian Grassmann integration associated with the $\psi$ field at inverse temperature $\beta$: 
\begin{equation} Q_\Lambda= {\rm Pf}C_m\,{\rm Pf}C_c\, e^{F_\Lambda(\lambda)}\int P(\mathcal D\psi) e^{\tilde V(\psi)}. \end{equation}
where $F_\Lambda(\lambda)$ is extensive in $\Lambda$, of order $\lambda$ and analytic in $\lambda$ in a small neighborhood of the origin (it is the $O(\lambda)$
contribution to the free energy from the integration of the massive degrees of freedom) and $\tilde V$ is a modified, effective, interaction that, similarly to the 
bare one, is of order $\lambda$, and it is the sum of Grassmann monomials in $\psi$ of order $2$, $4$, $6$, etc., whose kernels 
are analytic in $\lambda$ in a small neighborhood of the origin and 
decay exponentially to zero at large distances, with rate proportional to the inverse lattice spacing. 

\medskip

{\it 3. Setting up the multiscale integration: dressed reference Gaussian integration  and counterterms.} 
The idea now is to iterate the previous integration procedure. Of course, we cannot expect that a naive repetition of the 
strategy used to integrate the massive $\chi$ field out will work, due to the slow, polynomial, decay of the propagator of the $\psi$ field: if we tried to integrate the $\psi$ field out in a `single shot', as done for the $\chi$ field, we would get poor, non-uniform, estimates as $L,M\to\infty$ and/or $a\to 0$. On the contrary, our goal is to get estimates uniform in the scaling limit. For this purpose, as usual in cases of this sort, 
we use a multiscale procedure. First of all, recalling that the inverse covariance of the $\psi$ field is $\beta$-dependent, $C_c=C_c(\beta)$, we add and subtract at exponent a quadratic term $-\frac{Z}2(\psi,C_c^*\psi)$, where $C_c^*=C_c(\beta_c(0))$ is the  critical covariance, i.e., the one corresponding to the polynomially decaying propagator $g_c^*=(C_c^*)^{-1}$, and $Z$ is a multiplicative renormalization constant, to be fixed appropriately (a posteriori). 
Next, we rescale $\psi\to Z^{-1/2}\psi$, thus getting 
\begin{equation} Q_\Lambda= {\rm Pf}C_m\,{\rm Pf}C^*_c\, e^{F_\Lambda(\lambda)}Z^{|\Lambda|}\int P^*(\mathcal D\psi) e^{V^{(N)}(\psi)}, \label{eq:12}\end{equation}
where $P^*(\mathcal D\psi)$ is the Gaussian Grassmann integration with propagator $g_c^*$ and 
\begin{equation}V^{(N)}(\psi)=\tilde V(Z^{-1/2}\psi)+\frac12(\psi,(C_c^*-Z^{-1/2}C_c(\beta))\psi).\label{eq:13}\end{equation}
The term $\frac12(\psi,(C_c^*-Z^{-1/2}C_c(\beta))\psi)$ plays the role of a counterterm, with the temperature $\beta$ and the constant $Z$ to be fixed in such a way that the subsequent multiscale expansion is convergent, and the dressed propagator (i.e., the average of $\psi_{\omega,x}\psi_{\omega',y}$ with respect to the 
Grassmann `measure' in \eqref{eq:8}) is polynomially decaying, with the same asymptotic behavior as $\frac1{Z}g_c^*(x,y)$
at large distances. The resulting value of $\beta$ to be chosen so that these properties hold defines the interacting critical temperature $\beta_c(\lambda)$. 

The label $N$ in \eqref{eq:13} is $N=\lfloor \log_2a^{-1}\rfloor$ and has the meaning of (diadic) scale of the lattice spacing. The potential $V^{(N)}$ 
is called the {\it effective potential} on scale $N$. 

\medskip

{\it 4. Multiscale integration of the $\psi$ field.} We decompose the propagator $g_c^*$ associated with the reference Gaussian integration $P^*(\mathcal D\psi)$ in 
\eqref{eq:12} as follows: 
\begin{equation} g^*_c(x,y)=\sum_{h\le N} g^{(h)}(x,y),\end{equation}
where $g^{(h)}(x,y)$ has the following (approximate) scaling property\footnote{This scale covariance property is necessarily approximate in finite volume and at finite 
lattice spacing, but it becomes exact in the limit of infinite volume and lattice mesh to zero. Error terms are explicit and can be explicitly bounded, but we do not need to 
specify them for the purpose of the present discussion.}: \begin{equation}g^{(h)}(x,y)\simeq 2^h g^{(0)}(2^hx,2^hy),\end{equation} with $g^{(0)}$ an exponentially decaying propagator, with decay 
rate of order $1$. Correspondingly, using the addition property of Gaussian integrals, we rewrite $\psi$ as a sum of independent fields, $\psi=\sum_{h\le N}\psi^{(h)}$, 
where each $\psi^{(h)}$ is associated with a reference Gaussian integration $P_h(\mathcal D \psi^{(h)})$ with propagator $g^{(h)}$, thus getting: 
\begin{equation}\begin{split} Q_\Lambda&= e^{F_\Lambda^{(N)}}\int \prod_{h\le N}P_h(\mathcal D\psi^{(h)}) e^{V^{(N)}(\sum_{h\le N}\psi^{(h)})} \\
&\equiv e^{F_\Lambda^{(N)}}\int P_{\le N}(\mathcal D\psi^{(\le N)}) e^{V^{(N)}(\psi^{(\le N)})}, \label{eq:15}\end{split}\end{equation}
where $F_\Lambda^{(N)}=F_\Lambda(\lambda)+|\Lambda|\log Z+ \log {\rm Pf}C_m+ \log {\rm Pf}C_c^*$. 

The idea now is to perform the integration of the fluctuation fields $\psi^{(N)}$, $\psi^{(N-1)}$, etc, one at the time. Each step can be performed in full analogy with the 
integration of the massive field $\chi$, since the propagator of each field $\psi^{(h)}$ is exponentially decaying at large distances. Therefore,  after the integration of 
$\psi^{(N)}$, $\ldots$, $\psi^{(h+1)}$, we are left with 
\begin{equation} Q_\Lambda= e^{F_\Lambda^{(h)}}\int P_{\le h}(\mathcal D\psi^{(\le h)}) e^{V^{(h)}(\psi^{(\le h)})}, \label{eq:16}\end{equation}
where $F_\Lambda^{(h)}$ is analytic in $\lambda$ 
in a small neighborhood of the origin and $V^{(h)}$  is of order $\lambda$, and it is the sum of Grassmann monomials 
in $\psi$ of order $2$, $4$, $6$, etc., whose kernels are analytic in $\lambda$ in a small neighborhood of the origin and 
decay exponentially to zero at large distances, with rate proportional to $2^h$. 

Note, however, that the analyticity domain a priori may shrink step after step, as $N-h$ grows larger and larger. In order to prove uniform bounds on the radius of 
convergence we have to monitor the behavior of the kernels of the effective potential $V^{(h)}$ as $N-h$ grows: in particular we have to identify the terms that, on the 
basis of dimensional bounds, may grow under iterations; once these potentially dangerous terms have been identified, we need to look for cancellations in their 
perturbative expansions that may lead to improved bounds. 

The structure of the effective potential on scale $h$ is the following\footnote{To be precise, the correct structure of the effective potential is slightly more general than \eqref{eq:17}, in that derivative operators of order one or two on the Grassmann fields are allowed (therefore, \eqref{eq:17} should include an extra summation over an index $\bs D$, labelling how many derivatives are there, and which fields they act on): while the effective potential at the initial scale, $h=N$, is exactly of the form \eqref{eq:17}, the localization and interpolation procedure mentioned below generates derivatives acting on the Grassmann fields. These are crucial in order for the bounds on the kernels of the effective potential to be uniform in 
$L,M,a,h$.}:
\begin{equation} V^{(h)}(\psi)=\sum_{n\ge 2}\sum_{\bs \omega}\int d\bs x\,\psi(\bs \omega,\bs x)W_{n}^{(h)}(\bs \omega,\bs x),\label{eq:17}\end{equation}
where: the sum over $n$ runs over the even integers; for given $n$, the sum over $\bs \omega$ runs over $n$-ples of elements of $\{+,-\}$, 
and $\int d\bs x\equiv a^{2n}\sum_{\bs x}$ with $\sum_{\bs x}$ the sum over $n$-ples of points of $\Lambda$; for given $\bs \omega=(\omega_1,\ldots, \omega_n)$, 
and $\bs x=(x_1,\ldots,x_n)$, we let $\psi(\bs\omega,\bs x)=\psi_{\omega_1,x_1}\cdots \psi_{\omega_n,x_n}$; the kernel $W_{n}^{(h)}$ is anti-symmetric under simultaneous exchange of the elements of $\bs\omega$ and $\bs x$, and has the natural translation invariance properties associated with the boundary conditions 
under consideration (toroidal or cylindrical). As we shall see, $W_{n}^{(h)}$ decays exponentially at large distances, with rate proportional to $2^h$ (recall that $2^N\simeq a^{-1}$). For this reason, it is natural to rescale the argument $\bs x$ by a factor $2^{-h}$ in order to obtain a kernel that decays exponentially on a scale of order one. 
More precisely, we introduce the following `a-dimensional kernel':
\begin{equation} \mathcal W_n^{(h)}(\bs \omega,\bs x')= 2^{-2h(n-1)}2^{h(n/2-2)} W_n^{(h)}(\bs \omega,2^{-h}\bs x'),\end{equation}
where $\bs x'\in(2^h\Lambda)^n$. We shall measure the size of this rescaled kernel in terms of the following weighted $L_1$ norm: 
\begin{equation} \|\mathcal W_n^{(h)}\|_{\kappa;h}=\sup_{\bs\omega}\frac1{2^{2h}|\Lambda|}\int d\bs x' e^{\kappa \delta(\bs x')}\big|\mathcal W_n^{(h)}(\bs \omega,\bs x')\big|,\end{equation}
where: $\int d\bs x'$ is a shorthand for the Riemann sum $a^{2n}2^{2nh}\sum_{\bs x'\in(2^h\Lambda)^n}$;
$\delta(\bs x')$ is the tree distance among the elements of $\bs x'$, i.e., the Euclidean length of the shortest tree connecting them; and $\kappa$ is a sufficiently small 
constant, 
which can be chosen, e.g., to be half the rate of exponential decay of the propagator $g^{(0)}$. The iterative integration procedure sketched above naturally 
leads to the following bound: $\|\mathcal W_n^{(h)}\|_{\kappa;h}\le (C_h)^n|\lambda|^{\max\{1,\frac{n}{2}-1\}}$, 
for some positive constant $C_h$. The goal is to show that, by properly choosing $\beta$ and $Z$, the constant $C_h$ can be chosen to be independent of $h$, that is, 
that the bound on the rescaled, `a-dimensional', kernels can be improved to 
\begin{equation} \|\mathcal W_n^{(h)}\|_{\kappa;h}\le C^n\begin{cases} |\lambda|^{\frac{n}{2}-1} & n\ge 4\\
|\lambda| & n=2\end{cases}\label{eq:20}\end{equation}
for a suitable constant $C>0$. 

\medskip

{\it 5. The Wilsonian RG map: scaling dimensions and localization.}
The map
\begin{equation} W\!RG_h: \{\mathcal W_n^{(h)}\}_{n\ge 2} \to \{\mathcal W_n^{(h-1)}\}_{n\ge 2}\end{equation}
 from the collection of kernels on scale $h$ to those on scale $h-1$, consisting in the two steps `integrate out the degrees of freedom on scale $h$' $+$ `rescaling', 
 defines the {\it Wilsonian RG map} (the dependence of $W\!RG_h$ upon the scale is very weak, it is due just to finite size and finite lattice spacing effects, and it disappears
as $L,M\to\infty$ and $a\to 0$). In order to prove the bounds \eqref{eq:20}, it is appropriate to think the a-dimensional kernels $\mathcal W_n^{(h)}$ as being 
obtained from the effective potential at scale $N$ via the iterative application of $W\!RG_k$ on scales $k>h$, and to study in detail the action of $W\!RG_k$ on the kernels of different order, 
as well as the action of its linearization around a Gaussian fixed point. The basic `dimensional' estimates on $\mathcal W_n^{(h)}$ follow from the computation of the eigenvalues of the linearization of $W\!RG_k$, which can be easily shown to be $2^{2-n/2}$, with $n$ a positive even integer, and the 
eigenvectors consisting of Grassmann monomials of order $n$.
The exponent $2-n/2$ plays the role of {\it scaling dimension}: it is positive for 
$n=2$, negative for $n>4$ and zero for $n=4$. This indicates that thequadratic terms in the effective action generically\footnote{I.e., unless some cancellations
take place, possibly after the fine tuning of a few suitable parameters} tend to expand exponentially under iterations 
of the RG map (these are the {\it relevant} terms, in the RG jargon); terms of order $6$ or more (the {\it irrelevant} terms) tend to contract exponentially; while the quartic 
terms are neutral at linear order (the {\it marginal} terms). The full, non-linear, control of the relevant and marginal term requires a more detailed analysis thereof. 

The standard procedure to analyze these terms and identify, whenever possible, cancellations leading to their control, uniformly in the iteration step, is to extract the 
local part from the kernels (i.e., their `most divergent part') and to re-express the rest, the non-local contribution, in terms of an expression involving additional derivatives (the higher the number of derivatives, the better the behavior under the RG map: the scaling dimension of a kernel of order $n$ in the Grassmann fields with $p$ derivatives is $2-n/2-p$, which is negative for $n=2$ and $p\ge 2$, for $n=4$ and $p\ge 1$, and for $n\ge 6$ and $p\ge 0$). In order to illustrate the idea behind this procedure, consider the 
quartic contribution to the (unrescaled) effective action $V^{(h)}(\psi)$, which has the form
\begin{equation} V_4^{(h)}(\psi)=\sum_{\omega_1,\ldots,\omega_4}\int dx_1\cdots dx_4\, \psi_{\omega_1,x_1}\cdots \psi_{\omega_4,x_4} W^{(h)}_4(\bs \omega,\bs x),
\label{eq:21}\end{equation}
with $\bs \omega=(\omega_1,\ldots,\omega_4)$, $\bs x=(x_1,\ldots,x_4)$. The {\it local part} of this expression, denoted $\mathcal L V_4^{(h)}(\psi)$, 
is defined to be the one obtained by replacing 
the non-local monomial $\psi_{\omega_1,x_1}\cdots \psi_{\omega_4,x_4}$ by its local counterpart, in which the four Grassmann fields are computed all at the same point, say at $x_1$: 
\begin{eqnarray} \mathcal L V_4^{(h)}(\psi)&=&\sum_{\omega_1,\ldots,\omega_4}\int dx_1\cdots dx_4\, \psi_{\omega_1,x_1}\cdots \psi_{\omega_4,x_1} W^{(h)}_4(\bs \omega,\bs x)\nonumber\\
&=& \sum_{\omega_1,\ldots,\omega_4}\int dx_1 \psi_{\omega_1,x_1}\cdots \psi_{\omega_4,x_1} \mathcal L_0W^{(h)}_{4;\bs \omega}(x_1),\end{eqnarray}
with $ \mathcal L_0W^{(h)}_{4;\bs \omega}(x_1)=\int dx_2\cdots dx_4 W^{(h)}_4(\bs \omega,\bs x)$. Now note the {\it key cancellation}: 
\begin{equation} \boxed{\mathcal L V_4^{(h)}(\psi)=0}\end{equation}
simply because $\psi_{\omega_1,x_1}\cdots \psi_{\omega_4,x_1}\equiv 0$, by the anti-commutation rule of the Grassmann variables and the fact that 
the indices $\omega_1,\ldots,\omega_4$ cannot be all different among each other (because they take only two values, $+$ and $-$). The remainder term, denoted 
$\mathcal R V^{(h)}_4(\psi)$, involves a difference between the non-local monomial $\psi_{\omega_1,x_1}\cdots\psi_{\omega_4,x_4}$ and its local counterpart, 
which can be written in interpolated form as 
\begin{eqnarray} && \psi_{\omega_1,x_1}\cdots\psi_{\omega_4,x_4}-\psi_{\omega_1,x_1}\cdots\psi_{\omega_4,x_1}=\\
&&=(x_2-x_1)\cdot\int_0^1 ds  \psi_{\omega_1,x_1}\partial\psi_{\omega_2,x_2(s)}\,\psi_{\omega_3,x_3(s)}\psi_{\omega_4,x_4(s)}+\text{similar terms}
\nonumber\end{eqnarray}
with $x_j(s)=x_1+s(x_j-x_1)$. Correspondingly, after a change of variables (i.e., after renaming $x_j(s)$ as $z_j$) the non-local part of \eqref{eq:21} can be written as 
\begin{eqnarray} \mathcal R V^{(h)}_4(\psi)&=&\sum_{\bs\omega}\int d\bs z\, \psi_{\omega_1,z_1}\partial\psi_{\omega_2,z_2}\, 
\psi_{\omega_3,z_3}\psi_{\omega_4,z_4}\mathcal R W^{(h)}_{4;\bs \omega}(\bs z;2)\nonumber\\
&+& \text{similar terms},\end{eqnarray}
where $\mathcal R W^{(h)}_{4;\bs \omega}(\bs z;2)$ (the kernel of the non-local remainder, written via the interpolation procedure sketched above -- the label $2$ in parenthesis indicates that the derivative in the corresponding Grassmann monomial acts on the second field) 
has scaling dimension $2-n/2-p$ with $n=4$ and $p=1$, that is it is {\it irrelevant} with scaling dimension $-1$, and, therefore, its a-dimensional 
counterpart shrinks exponentially under iterations of the RG map. The extraction of the local part of the quadratic contribution to $V^{(h)}(\psi)$, 
\begin{equation} V_2^{(h)}(\psi)= \sum_{\omega_1,\omega_2}\int dx_1\, dx_2\, \psi_{\omega_1,x_1} \psi_{\omega_2,x_2} W^{(h)}_2(\bs \omega,\bs x),
\label{eq:26}\end{equation}
proceeds analogously, the main difference being that $\psi_{\omega_2,x_2}$ is not replaced simply by $\psi_{\omega_2,x_1}$, 
but rather by $\psi_{\omega_2,x_1}+(x_2-x_1)\cdot\partial\psi_{\omega_2,x_1}$, i.e., by its Taylor expansion at $x_1$ of order $1$ (the criterium for stopping the expansion 
at order $1$ rather than $0$ being that the scaling dimension of the remainder term must be negative - and wouldn't have been so otherwise). Therefore, 
\begin{eqnarray} \label{eq:27}\mathcal L V_2^{(h)}(\psi)&=&\sum_{\omega_1,\omega_2}\int dx_1\Big( \psi_{\omega_1,x_1}\psi_{\omega_2,x_1} \mathcal L_0W^{(h)}_{2;\bs \omega}(x_1)\\ && \hskip1.8truecm+\psi_{\omega_1,x_1}\partial \psi_{\omega_2,x_1} \mathcal L_1W^{(h)}_{2;\bs \omega}(x_1)\Big),\nonumber\end{eqnarray}
with $\mathcal L_0W^{(h)}_{2;\bs \omega}(x_1)=\int dx_2 W^{(h)}_2(\bs \omega,\bs x)$ and 
$\mathcal L_1W^{(h)}_{2;\bs \omega}(x_1)=\int dx_2 (x_2-x_1)\cdot$ $\cdot W^{(h)}_2(\bs \omega,\bs x)$, while the corresponding remainder can be schematically 
written as 
\begin{equation} \mathcal RV_2^{(h)}(\psi)=\sum_{\omega_1,\omega_2}\int dz_1\,dz_2\, \psi_{\omega_1,z_1}\partial^2\psi_{\omega_2,z_2}\mathcal R W^{(h)}_{2;\bs \omega}(\bs z),\end{equation}
where $\mathcal R W^{(h)}_{2;\bs \omega}(\bs z)$
has scaling dimension $2-n/2-p$ with $n=2$ and $p=2$, i.e., it is {\it irrelevant} with scaling dimension $-1$, and, therefore, its a-dimensional 
counterpart shrinks exponentially under iterations of the RG map. 

Contrary to the local quartic part, the two terms in the right side of \eqref{eq:27} have no reason to cancel. In order to control their behavior under iterations of the RG, 
we need to exhibit cancellations, as discussed in the next item. 

\medskip

{\it 6. Flow of the effective coupling constants.} The specific structure of the local quadratic terms in the right side of \eqref{eq:27} depends on the boundary conditions 
chosen. 

In the case that $\Lambda$ is a torus, we have full translational invariance and, therefore, both $\mathcal L_0W^{(h)}_{2;\bs \omega}(x_1)$ 
and $\mathcal L_0W^{(h)}_{2;\bs \omega}(x_1)$ are independent of $x_1$.
Therefore, in this case, using also the underlying symmetries of the model under flip of the $\omega$ index, 
the local quadratic terms of the effective action takes the form 
\begin{equation} \int dx \Big( 2^h\nu_h\psi_{+,x}\psi_{-,x}+\zeta_h \sum_{\omega}\psi_{\omega,x} (\partial_1+i\omega\partial_2)\psi_{\omega,x}\Big),
\label{eq:29}\end{equation}
for two suitable constants $\nu_h$ and $\zeta_h$ (the $2^h$ in front of $\nu_h$ is chosen in such a way that the a-dimensional counterpart of that term has no 
$h$-dependent pre-factor in front). For generic initial data at scale $N$, the flows of $\nu_h$ and $\zeta_h$ tend to diverge exponentialy and logarithmically, respectively. 
However, it is easy to see (via an instance of the central manifold theorem or, equivalently, of the implicit function theorem) that it is possible 
to fine-tune the initial data $\nu_N$ and $\zeta_N$ in such a way that $\nu_h,\zeta_h$ remain bounded (and, actually, go to zero, in the thermodynamic and $a\to0$ 
limits). Remarkably, the two terms in \eqref{eq:29} have the same structure as the corresponding local terms of 
$\frac12(\psi,(C_c^*-Z^{-1/2}C_c(\beta))\psi)$, see \eqref{eq:13}, and it is possible to choose $\beta$ and $Z$ in such a way that the initial data $\nu_N,\zeta_N$ are the 
`right ones', i.e., those guaranteeing boundedness of $\nu_h,\zeta_h$, uniformly in $h,L,M,a$. In particular, the resulting choice of $\beta$ corresponds to $\beta=\beta_c(\lambda)$, the interacting critical temperature (the resulting choice of $Z$ corresponds, instead, to the multiplicative renormalization of the dressed 
fermionic propagator). This concludes the sketch of the proof in the translationally invariant setting. 

In the case that $\Lambda$ is a cylinder, the local contributions $\mathcal L_0W^{(h)}_{2;\bs \omega}(x_1)$ 
and $\mathcal L_0W^{(h)}_{2;\bs \omega}(x_1)$ explicitly depend on the vertical coordinate of $x_1$ (dependence on the horizontal coordinate disappears by translational invariance in the horizontal direction): therefore, the analogue of \eqref{eq:29} now reads 
\begin{equation}  \int dx \Big( 2^h\nu_h((x)_2)\psi_{+,x}\psi_{-,x}+\zeta_h((x)_2)\sum_{\omega}\psi_{\omega,x} (\partial_1+i\omega\partial_2)\psi_{\omega,x}\Big),
\label{eq:30}\end{equation}
where $(x)_2$ is the vertical coordinate of $x$.
We now add and subtract from $\nu_h((x)_2)$ its bulk counterpart, i.e., the coupling constant $\nu_h$ computed in the presence of periodic boundary 
conditions in both coordinate directions, and similarly for $\zeta_h((x)_2)$. The differences $\nu_h((x)_2)-\nu_h$ and $\zeta_h((x)_2)-\zeta_h$ decay to zero exponentially away from the boundary, 
i.e., they can be bounded proportionally to $e^{-\kappa 2^h{\rm dist}_2(x,\partial\Lambda)}$ for a suitable constant $\kappa>0$, with dist$_2$ the distance in the vertical 
direction, and $\partial\Lambda$ the (horizontal) boundary of $\Lambda$. {\it Key fact}: this additional 
exponential decay implies that the effective scaling dimensions of these boundary corrections is better by one scaling dimension than their bulk counterparts: 
therefore, the boundary correction to $\nu_h$ is dimensionally marginal, while the one to $\zeta_h$ is dimensionally irrelevant. The problem is thus reduced to 
the study of the marginal boundary correction to $\nu_h$. The idea here is to localize this term on the boundary, up to an additional remainder that is now 
dimensionally irrelevant: 
\begin{equation} \int dx\, 2^h(\nu_h((x)_2)-\nu_h)\psi_{+,x}\psi_{-,x}=
\nu^E_h \int_{\partial\Lambda} \!\!\!dx\, \psi_{+,x}\psi_{-,x}+\text{irrelevant remainder},\end{equation}
where the first term in the right side is the local edge term, dimensionally marginal. Remarkably, thanks to an exact cancellation 
of the propagator on the cylinder, related to an approximate image rule it satisfies, such local edge term is identically zero, simply because $\psi_{+,x}$, resp. $\psi_{-,x}$, vanishes on the bottom, resp. top, boundary of the cylinder. This allows us to fully control the flow of the 
local quadratic terms in the cylinder case and concludes our sketch of the proof. 

\section{Conclusions}

After a general introduction to conformal invariance of critical statistical mechanics models, in connection with the Wilsonian RG picture, 
I reviewed the recent results obtained via constructive RG methods on the scaling limit of non-integrable perturbations of the 2D Ising 
model \cite{AGG20,GGM12}. 
The stated results provide the first construction of the scaling limit of the model (or, better, of its energy sector) in the full plane and in the finite cylinder. 
The generalization from the full plane to the cylinder requires to introduce new ideas, regarding, in particular, the dimensional estimates of the edge terms, 
their localization and partial cancellation properties. The results and underlying proofs motivate a number of natural questions on Ising-type and related 
two-dimensional critical systems, which I hope will be addressed in the next future, also thanks to the collaborations stimulated by today's workshop: 
\begin{itemize}
\item How can we control the scaling limit in more general domains? The missing ingredient is a better control of the fermionic propagator of the nearest neighbor model in domains of arbitrary shape, not relying on its exact diagonalization. Note that the proofs by Smirnov, Chelkak, Hongler and Izyurov on the conformal invariance 
of the scaling limit of the nearest neighbor Ising model imply, in particular, that the fermionic propagator $g_a^\Omega$ at finite lattice spacing $a$ in an arbitrary 
domain $\Omega$ equals its scaling limit $g_0^\Omega$ plus a remainder that goes to zero as $a\to0$, with explicit estimates on the speed of convergence. 
The output of the works \cite{CHI15,CS09,HS13} may serve as an input of a generalized RG construction in more general domains. Unfortunately, the currently available bounds on 
the remainder $g_a^\Omega-g_0^\Omega$ are too weak for the RG machinery to start. However, it is `clear' that further progress on this topic will come from 
a combination of the methods of constructive RG with those of discrete holomorphicity: several discussions on this problem took place during the workshop
with D. Chelkak and H. Duminil-Copin, among others, and they defined a clear strategy to attack the technical questions involved. 
\item How can we control the scaling limit of the spin-spin correlations in the non-integrable perturbed case? A good starting point seems to be the representation 
of the spin-spin correlation in terms of a fermionic propagator on a discretized Riemann surface with a monodromy cut \cite{CHI15}. The effect of the cut has some (superficial) similarities with the effect of a boundary, for instance because it breaks translational invariance by the insertion of a one-dimensional defect. It is likely that progress on the problem of spin-spin correlations will emerge from a better understanding of how to implement constructive RG in non-translationally-invariant situations. 
\item How can we compute boundary effects in `marginal theories' such as interacting dimers \cite{GMT17,GMT20}? It is likely that the scheme sketched above, involving a decomposition of the effective couplings into a local bulk part $+$ local edge part $+$ irrelevant remainder, generalize straightforwardly to many other 2D models close to the 
free Fermi point, such as $XXZ$ spin chains or interacting dimers. In general, I don't expect that the local edge couplings will cancel for simple symmetry reasons: 
presumably, their flow will be non trivial and will be associated to an anomalous critical exponent, related to new anomalous critical exponents of the
boundary correlation functions. I hope to report new results on this exciting open problem in a future publication
\end{itemize}

\bigskip

{\bf Acknowledgements.} This work has been supported by the European Research Council (ERC) under the European Union's Horizon 2020 research and innovation programme ERC CoG UniCoSM, grant agreement n.724939. I also acknowledge financial support from MIUR, PRIN 2017 project MaQuMA, PRIN201719VMAST01.
I warmly thank the organizers of the workshop series {\it Inhomogeneous Random Systems} (IRS), Fran\c{c}ois Dunlop and Ellen Saada, for the opportunity they gave me to serve as moderator of the one-day workshop on {\it Emergent CFTs in statistical mechanics} and for the invaluable service they make to the statistical mechanics community (interpreted in a very broad sense) in keeping the IRS tradition alive year after year, thus stimulating a constant dialog among different areas and sub-communities.

\bibliographystyle{ws-procs975x65}

\end{document}